\documentclass[12pt,draftcls,onecolumn]{IEEEtran}
\ifCLASSINFOpdf
\else
\fi
\usepackage{epsfig}
\usepackage{color}
\usepackage{amssymb}
\usepackage{amsmath}
\usepackage{amsfonts}

\newtheorem{theorem}{Theorem}

\newtheorem{definition}{Definition}
\newtheorem{lemma}{Lemma}
\newtheorem{corollary}{Corollary}

\newtheorem{remark}{Remark}

\begin{document}
%

\title{Game-based coalescence over multi-agent systems}
%
%
%

\author{Jingying~Ma,~
       Jinming  Du,~
        and Yuanshi~Zheng
\thanks{This research was supported by the the National Natural Science Foundation of China (Grant Nos. 61963032, 61703082, 71790614, 71621061, 71520107004, 61773303 and 61751301), the 111 Project (B16009), and Natural Science Foundation of Ningxia (Grant No. 2018AAC03033).}
\thanks{J. Ma is with School of Mathematics and Statistics, Ningxia University, Yinchuan 750021, P.~R.~China
 (e-mail: majy1980@126.com)}
 \thanks{J. Du is with Institute of Industrial and Systems Engineering, College of Information Science and Engineering, Northeastern University, Shenyang 110819, China, Liaoning Engineering Laboratory of Operations Analytics and Optimization for Smart Industry, Northeastern University, Shenyang 110819, P.~R.~China, and Key Laboratory of Data Analytics and Optimization for Smart Industry (Northeastern University), Ministry of Education, Shenyang 110819, P.~R.~China
 (e-mail: dujinming@ise.neu.edu.cn)}
\thanks{Y. Zheng (Corresponding author) is with the Center for Complex Systems, School of Mechano-electronic Engineering, Xidian University, Xi'an 710071, P.~R.~China (e-mail: zhengyuanshi2005@163.com)
 }
}

\maketitle

\begin{abstract}
Coalescence,  as a kind of ubiquitous group behavior in the nature and society, means that agents, companies or other substances keep consensus in states and act as a whole. This paper considers coalescence for $n$ rational agents with distinct initial states. Considering the rationality and intellectuality of the population,  the coalescing process is described by a bimatrix game which has the unique mixed strategy Nash equilibrium solution. Since the process is not an independent stochastic process, it is difficult to analyze the coalescing process. By using the first Borel-Cantelli Lemma, we prove that all agents will coalesce into one group with probability one. Moreover, the expected coalescence time is also evaluated. For the scenario where payoff functions are power functions, we obtain the distribution and expected value of coalescence time. Finally, simulation examples are provided to validate the effectiveness of the theoretical results.
\end{abstract}



%
\IEEEpeerreviewmaketitle

\section{Introduction}\label{s1}

Recently, group behaviors of individuals have attracted the attention of many disciplines, such as sociology\cite{Hegselmann2002Opinion}, economics\cite{Bullo2017Competitive,du2019evolutionary}, biology\cite{kao2014collective} and engineering\cite{ren2007distributed}.  Roughly speaking, group behaviors of multiple agents include consensus\cite{ren2005consensus}, flocking\cite{Morin2015Collective}, containment\cite{zheng2014containment}, leader emergence\cite{Couzin2005Effective,Pais2014Adaptive} and so on. Among above aspects of group behavior, how to make a group of individuals reaching consensus is a fundamental and important issue.  Consensus means that agents reach an agreement upon certain quantities of interest, such as opinions of social individuals\cite{Hegselmann2002Opinion,Claudio2012dynamics}, and speeds of mobile autonomous robots\cite{Jadbabaie2003}. Many results were obtained, to name but a few, leader-following consensus\cite{Majingying2015LQR}, consensus problems for multiple double-integrator agents\cite{xie2007consensus,ren2008consensus} and for agents with different dynamics\cite{Qin2017On,Zheng2017Consensus,ma2019consensus}.

The above literature mostly assumed that each agent is simple, and thereby just obeies a uniform rule without figuring out its own interest. However, in the real world, one noteworthy feature is that agents are diverse. For example, they have different objectives or interests. Another prominent feature of agents is of high intelligence --- they choose the best possible response based on their interests. Thus, the relationships among agents might be noncooperative, even competitive, and the interaction of them might be playing games instead of obeying the fixed protocols. Based on game theory, some complex group behavior, such as competitive propagation\cite{Bullo2017Competitive,ma2016Equilibrium}, network formation\cite{jackson2002evolution}, collective learning\cite{kao2014collective} and coalescence \cite{cox1989coalescing}, were studied. To achieve a global task, agents need to coalesce, i.e., to form a group where they can make decisions together and act as a whole. Coalescence is common seen in real world, such as coalescence for robot groups \cite{poduri2007latency} and coalescence of opinions in social networks\cite{cox1989coalescing,cooper2013coalescing}.

Inspired by the above references, we consider coalescence of $n$ agents with distinct initial states. It means that agents will finally keep consensus in states and act as a whole. It is necessary to mention the difference between consensus and coalescence. Consensus means agents' states reach or asymptotically converge to an identical value. Whereas, coalescence is more complicated than consensus. To reach coalescence, agents need to reach an agreement on states in the finite time, and from then on alway keep consensus not only in states but also in action. Therefore, the essential question we face is, how to design a mechanism to make agents coalesce into a group. We assume that each agent is rational and accesses complete information, i.e.,  each agent chooses the best response based on its interest and the global information of the population. Based on this assumption, we propose a kind of bimatrix games where each player has two strategies to choose -- cooperation (C) and defection (D). Cooperation means players sacrifice part of interests and change their states to achieve coalescence. On the contrary, defection means players tend to keep their states regardless of whether coalescing or not. By playing this game, agents coalesce into groups, then the agents in the same group act as a whole and play games with those in other groups. By merging groups and groups, they eventually coalesce into one group. We find that the game has the unique mixed strategy Nash equilibrium --- players choose strategies in a probabilistic sense, which makes the coalescing process be a stochastic process. Because it depends on payoff functions,  it is not an independent stochastic process. As a result, it is not easy to analyze the coalescence of the population.
The contributions of this paper are summarized as follows.
\begin{itemize}
\item We establish a kind of bimatrix game model to show the interaction among agents. We prove that the game has the unique mixed strategy Nash equilibrium solution.
\item By virtue of the first Borel-Cantelli Lemma, we prove that all the agents coalesce into one group with probability one.
\item The distribution and the expected of coalescence time are evaluated.
\end{itemize}

%
%

The rest of this paper is organized as follows. In Section \ref{s2}, we introduce some basic notions of bimatrix game. Section \ref{s3} shows our main results. Numerical simulations are given in Section \ref{s4} to illustrate the effectiveness of theoretical results. Some conclusions are drawn in Section \ref{s5}.

Throughout this paper, the following notations will be used: let $\mathbb{R}$, $\mathbb{R}_{\geq 0}$ be the sets of real numbers and  nonnegative real numbers, respectively. $\mathbb{R}^{n\times m}$ is the set of $n\times m$ real matrices. $\mathcal{I}_n=\{1, \cdots, n\}$ is an index set.  For a random event $A$, $\mathbb P(A)$ means the probability of event $A$. For a  random variable $S$, $\mathbb E(S)$ and $\mathbb D (S)$ mean the expected value and the variance of $S$ respectively.


\section{Preliminaries}\label{s2}

\subsection{A brief introduction for bimatrix games}
In this subsection, we introduce some basic notions about bimatrix game. For more details, interested readers are referred to \cite{Ba1999Dynamic}.

Suppose that two players $P_1$ and $P_2$ play a game.  $P_1$ has strategies $r_1, r_2, \cdots, r_m$, and $P_2$ has strategies $c_1, c_2, \cdots, c_n$. If $P_1$ adopts the strategy $r_i$ and $P_2$ adopts the strategy $c_j$, then $(r_i,c_j)$ is a pair of pure strategies, and $a_{ij}$ (respectively, $b_{ij}$) denotes the profit incurred to $P_1$ (respectively, $P_2$). Each player seeks to maximum its own profit by independent and simultaneous decision. This game is comprised of two $(m \times n)$-dimensional matrices, $A =\{a_{ij}\}$ and $B = \{b_{ij}\}$, with each pair of entries $(a_{ij},b_{ij})$ denoting the payoff of the game corresponding to a particular pair of decisions made by the players. Thus, this game is called the bimatrix game $(A, B)$.
A pair of strategies $\{r_{i^*}, c_{j^*}\}$ is said to constitute a pure strategy Nash equilibrium solution to a bimatrix game $(A, B)$ if the following pair of inequalities is satisfied for all $i \in \mathcal I _{m}$, $j \in \mathcal I _{n}$:
$\left\{
\begin{aligned}
& a_{i^*j^*}\geq a_{ij^*} ,\\
& b_{i^*j^*}\geq b_{i^*j}. \\
\end{aligned}
\right.
$
Furthermore, the pair $(a_{i^*j^*},b_{i^*j^*})$ is known as a pure strategy Nash equilibrium of the bimatrix game.  In many cases, pure strategy Nash equilibrium strategies might not exist. Hence, we now enlarge the concepts of strategy and Nash equilibrium, which are defined as the set of all probability distributions on the set of pure strategies of each player.
We call $\Gamma _1=\{r_1,r_2, \cdots, r_m\}$ and $\Gamma _2=\{c_1,c_2, \cdots, c_n\}$ are the strategy spaces of players $P_1$ and $P_2$, respectively.   Let $\alpha=[\alpha_1,\cdots,\alpha_m]^T$ be a non-negative vector satisfying $\sum_{i=1}^m\alpha_i=1$, where $\alpha_i$ denotes player $P_1$ will choose strategy $r_i$ with probability  $\alpha_i$. Obviously, $\alpha$ is the probability distribution of the strategy space $\Gamma_1$. We define that $\alpha$ is a mixed strategy of $P_1$. Likewise, $\beta=[\beta_1,\cdots, \beta_n]^T$ is a mixed strategy of $P_2$.  Suppose that the game is played repeatedly, and the outcomes which are maximized by players is determined by averaging the outcomes of the player. Hence, we call $(\alpha,\beta)$ as a pair of mixed strategies, and $U_1(\alpha,\beta)=\alpha^TA\beta$ and $U_2=(\alpha,\beta)=\alpha^TB\beta$ as the corresponding utilities of $P_1$ and $P_2$, respectively. Each player decides its mixed strategy independently to maximize its utility.  Subsequently, we give the definition of mixed strategy Nash equilibrium\cite{Ba1999Dynamic}.


\begin{definition}
A mixed strategy pair $\{\alpha^*, \beta^*\} $ is said to constitute a mixed strategy Nash equilibrium solution to a bimatrix game $(A, B)$, if
the following inequalities are satisfied for all mixed strategy pairs:
\[
\begin{aligned}
 (\alpha^*)^TA\beta^* \geq \alpha^TA\beta^* ,\\
  (\alpha^*)^TB\beta^{*} \geq (\alpha^*)^TB\beta .
\end{aligned}
\]
\end{definition}

%



\subsection{The first Borel-Cantelli lemma}

At the end of this section, we introduce the first Borel-Cantelli lemma which will be used in our paper.

\begin{lemma}\label{Borel¨CCantelli}(The first Borel-Cantelli lemma\cite{Gut2005Probability}
)Let $\{A_k, k \geq 1\}$ be arbitrary events. If the sum of the probabilities of the $A_{k}$ is finite, then the probability that infinitely many of them occur is 0, that is,
$
\mathbb P(\cup_{l=1}^{\infty} \cup_{k\geq l}^{\infty} A_k)=0.
$
\end{lemma}





\section{\bf Coalescence of multiple agents}\label{s3}

Consider a system with $n$  agents labeled $1,2,\cdots,n$ where each agent $i$ ($i\in \mathcal I_n$) has the state $x_i(k)\in \mathbb R^{m}$ at time $k=0,1,\cdots$. Throughout this paper, we assume that

 Assumption 1: All agents are rational and complete information accessible.

Assumption 2: At time $k=0$, each agent composes one group and has a distinctive state, i.e.,  $x_i(0) \neq x_j(0)$ for all $i \neq j$.

Assumption 3: Agents who are in the same group will make decisions together, share information simultaneously and keep consensus on states.

In this paper, we consider how to make $n$ agents coalescing, i.e., merging into one group where they can make decisions together, share information simultaneously and keep consensus on states. We first propose the notion of coalescence for the system.

\begin{definition}\label{Def_consensus}
For a multi-agent system composed of $n$  agents with distinct initial states, if there exists a minimum time $K^{*}$ such that, starting from time $K^*$,  all agents make decisions together, share information simultaneously and keep consensus on states, then the system is said to reach coalescence at time $K^{*}$. Random variable $K^{*}$ is called the coalescence time of the system. $\mathbb E(K^{*})$ is called the expected coalescence time.
\end{definition}

%

\subsection{The interaction among groups }

In this subsection, we propose a bimatrix game $\mathbb G$ to model the interaction of groups. Moreover, the unique mixed strategy Nash equilibrium solution of game $\mathbb G$ is obtained.

\it{Players:} There are two players $P_1$ and $P_{2}$. Players decide whether or not to change their states by playing games.  Let the states of $P_1$ and $P_{2}$ before game be $y_{1}$ and $y_{2}$ and after game be $y_{1}'$ and $y_{2}'$ respectively ($y_{r}, y_{r}'\in \mathbb R^{m}, r=1,2$).

\begin{figure}
  \centering
  \includegraphics[width=6cm]{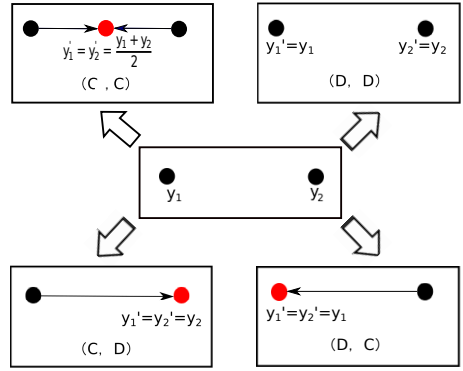}\\
  \caption{Four strategy pairs and out-comings of Game $\mathbb G$}\label{fig0}
  \end{figure}

 \it{Strategies:}
 Each player has two strategies to choose from---  cooperation ($C$) and defection ($D$). If a player chooses $C$, it means this player will change its state to coalesce with the other player.  If a player chooses $D$, this agent will not change its state regardless of whether they can coalesce or not. Therefore,  there are four strategy pairs and the corresponding out-comings (presented in Table \ref{talbel1} and Fig. \ref{fig0}):
 \begin{itemize}
 \item If both two players choose $C$, i.e., the strategy pair $(C, C)$, both of them will update states to the middle of their states to coalesce into a group;
 \item If one player chooses $C$ and the other chooses $D$,  i.e., the strategy pair $(C, D)$ or $(D, C)$, only the cooperative player will change its state to that of the other one's and they coalesce into a group;
 \item If both two players choose $D$, i.e., the strategy pair $(D, D)$,  no one will change its state and thereby merging fails.
 \end{itemize}

 \it{Payoff:}
Each player will face two kinds of interests--- cost of state changing and profit of coalescence. For player $P_{r}(r=1,2)$, the cost of state changing is $f(||y_r'-y_r||)$,  and the profit of coalescence is $g(||y_1-y_2||)$, where $f: \mathbb R_{\geq 0} \mapsto \mathbb R_{\geq 0}$ and $g: \mathbb R_{\geq 0} \mapsto \mathbb R_{\geq 0}$ are strictly monotone increasing continuous functions with $f(0)=g(0)=0$.  Therefore, the payoff of player $P_r$ is $g(||y_1-y_2||)-f(||y_r'-y_r||)$.

Suppose that two players choose strategies independently and simultaneously. Let $\xi=||y_{1}-y_{2}||$. The strategy pairs, outcomes, and payoffs of the game are listed in Table  \ref{talbel1}.

\begin{table}[!hbt]
  \centering
  \caption{The outcomes and payoffs of the game $\mathbb G$.}
  \label{talbel1}
\begin{tabular}{c|c|c|c}
\hline  
$(s_1,s_2)$ & Outcomes of states & Payoff of $P_{1}$ & Payoff of $P_{2}$\\
\hline  
$(C,C)$ & $y_{1}'=y_{2}'=\frac{y_{1}+y_{2}}{2}$ & $g(\xi)-f(\frac{\xi}{2})$ & $g(\xi)-f(\frac{\xi}{2})$ \\
$(C,D)$ & $y_{1}'=y_{2}'=y_{2}$ & $g(\xi)-f(\xi)$ & $g(\xi)$ \\
$(D,C)$ & $y_{1}'=y_{2}'=y_{1}$ & $g(\xi)$ & $g(\xi)-f(\xi)$ \\
$(D,D)$ & $ y_{1}'=y_{1},y_{2}'=y_{2}$ & $0$ & $0$ \\
\hline 
\end{tabular}
\end{table}

\begin{remark}
 Easy to find that the game will represent a prisoner's dilemma if $f(\xi)>g(\xi)$, i.e., each player will choose $D$, which means that all agents always keep their initial states. Therefore, in the remaining parts of the paper we assume that $f(\xi)<g(\xi)$.
\end{remark}

\begin{theorem}
Game $\mathbb G$ has the unique mixed strategy Nash equilibrium solution, that is each player choosing $C$ with probability $\frac{f(\xi)-g(\xi)}{g(\xi)-f(\xi)+f(\frac{\xi}{2})}$ and $D$ with probability  $\frac{f(\frac{\xi}{2})}{g(\xi)-f(\xi)+f(\frac{\xi}{2})}$.
\end{theorem}
{\it Proof.} According to the definition of bimatrix game, game $\mathbb G$ is a bimatrix game where $\Gamma_1=\Gamma_2=\{C,D\}$ and
\[
A=\begin{pmatrix}
g(\xi)-f(\frac{\xi}{2}) & g(\xi)-f(\xi)\\
g(\xi) & 0
\end{pmatrix},
B=A^T.
\]
Let $\alpha=[p,1-p]^T$ and $\beta=[q,1-q]^T$ be the mixed strategies of $P_{1}$ and $P_{2}$, respectively. Then, the utilities of $P_{1}$ and $P_{2}$ are
$U_1(p,q)=[p,1-p]A[q,1-q]^T$ and $U_2(p,q)=[p,1-p]B[q,1-q]^T.$

Suppose that $\{[p^*,1-p^*]^T,[q^*,1-q^*]^T\}$ is the mixed strategy Nash equilibrium solution of the game $\mathbb G$.
By the definition of mixed strategy Nash equilibrium, we have
$$
\left\{
\begin{aligned}
 &U_1(p^*,q^*) \geq U_1(p,q^*)\\
 &U_2(p^*,q^*) \geq U_2(p^*,q)\\
 \end{aligned}
\right.,
$$
which means that
\begin{equation}\label{Nash-c}
\left\{
\begin{aligned}
&\left.\frac{\partial U_1(p,q^*)}{\partial p}\right\vert_{p=p^*}=0,\\
&\left.\frac{\partial U_2(p^*,q)}{\partial q}\right\vert_{q=q^*}=0.\\
\end{aligned}
\right.
\end{equation}
By solving (\ref{Nash-c}), we have $p^*=q^*=\frac{g(\xi)-f(\xi)}{g(\xi)-f(\xi)+f(\frac{\xi}{2})}$. Hence, we know that the Nash equilibrium solution in the mixed strategies is that  each player chooses $C$ with probability $\frac{g(\xi)-f(\xi)}{g(\xi)-f(\xi)+f(\frac{\xi}{2})}$ and $D$ with probability  $\frac{f(\frac{\xi}{2})}{g(\xi)-f(\xi)+f(\frac{\xi}{2})}$. ~$\blacksquare$

\begin{corollary}\label{cor1}
Two players will coalesce into one bigger group with probability $1-\left(\frac{f(\frac{\xi}{2})}{g(\xi)-f(\xi)+f(\frac{\xi}{2})}\right)^2$.
\end{corollary}
{\it Proof.} By the definition of the game, we know that two players will coalesce if and only if $y_{1}'=y_{2}'.$ It is easy to find from Table 1 that
\begin{equation*}\label{p-C}
\mathbb P(y_{1}'=y_{2}')=1-\left(\frac{f(\frac{\xi}{2})}{g(\xi)-f(\xi)+f(\frac{\xi}{2})}\right)^2 ~.~\blacksquare
\end{equation*}

\begin{figure}
  \centering
  \includegraphics[width=9cm]{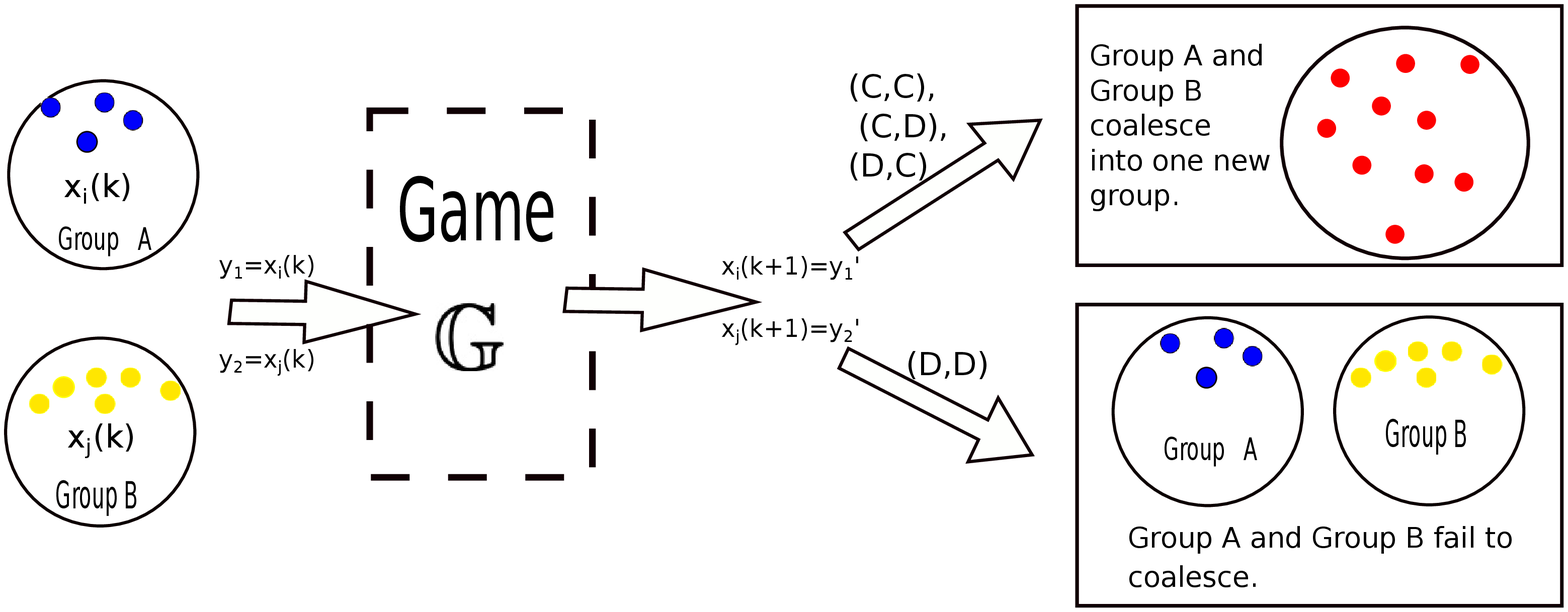}\\
  \caption{the dynamic of the system}\label{fig4}
  \end{figure}

The interaction among groups can be described in the following manner (See Fig. \ref{fig4}): at each time $k$, two groups are chosen to play the game $\mathbb G$, where all involved  agents will update their states according to the rules of game $\mathbb G$. If merger occurs in the game  $\mathbb G$, two groups will coalesce into one group.

\begin{remark}
Since each member of a group has the same state, they have the same interest. Therefore, we assume that strategy selection is determined by all agents of the group. Agents, who make decisions together, obtain the identical payoff simultaneously. Suppose that two groups consist of $s_{1}$ agents and $s_{2}$ agents respectively. Define
\[
 U^{(k)}(p^{*},q^{*})=s_{1}U_{1}(p^{*},q^{*})+s_{2}U_{2}(p^{*},q^{*})
\]
as the aggregate expectational payoff of two groups at time $k$.
\end{remark}


\subsection{ Coalescence of multiple agents: general cases}

Let $\xi_{k}=|y_{1}(k)-y_{2}(k)|$ where $y_{1}(k)$ and $y_{2}(k)$ represent the states of two players before playing game $\mathbb G$ at time $k$. Then, we can calculate
\[U^{(k)}(p^{*},q^{*})=(s_{1}+s_{2})\frac{(g(\xi_{k})-f(\xi_{k}))g(\xi_{k})}{g(\xi_{k})-f(\xi_{k})+f(\frac{\xi_{k}}{2})}.
\]
Let $p_{k}$ indicate the probability of ``two players coalesce  at time $k$''.  Then, we have $p_{k}=1-\left(\frac{f(\frac{\xi_{k}}{2})}{g(\xi_{k})-f(\xi_{k})+f(\frac{\xi_{k}}{2})}\right)^2.$

\begin{lemma}\label{thm:p_low_and_pup}
For the initial states $x_{1}(0),\dots,x_{n}(0)$, there exist two positive constants $0<p_{low}<p_{up}<1$  such that
$p_{low}\leq p_{k}\leq p_{up}$ for all $k=1,2,\dots, K^{*}$.
\end{lemma}
{\it Proof.}   For the initial states $x_{1}(0),\dots,x_{n}(0)$, we have
\[
\xi_{min}\leq\xi_{k}\leq \xi_{max}.
\]
where
$
\xi_{min}=\min_{i,j\in \{1,2,\dots,n\}}|x_{i}(0)-x_{j}(0)|
$
and
\[\xi_{max}=\max_{i,j\in \{1,2,\dots,n\}}|x_{i}(0)-x_{j}(0)|.\]
Since $f(.)$ and $g(.)$ are continuous, $\frac{f(\frac{\xi}{2})}{g(\xi)-f(\xi)+f(\frac{\xi}{2})}$ is also continuous in $[\xi_{min},\xi_{max}]$. It is easy to obtain that,  there exist two positive constants $0<\nu< \mu<1$  such that
$0<\nu \leq \frac{f(\frac{\xi}{2})}{g(\xi)-f(\xi)+f(\frac{\xi}{2})} \leq \mu<1$ for all $\xi_{k}\in [\xi_{min},\xi_{max}]$. Thus, we have
$1-\mu^{2}=p_{low}<p_{k}<p_{up}=1-\nu^{2}.~\blacksquare$


\begin{theorem}\label{thm:est-PK}
If Assumptions 1 - 3 hold and all groups interact by playing game $\mathbb G$, then the probability with which the coalescence time equals to $T(\geq n-1)$ can be estimated  by
\begin{equation}\label{eq:estimation-of-K}
\begin{aligned}
C_{T-1}^{n-2}(1-p_{up})^{T+1-n}p_{low}^{n-1}\leq \mathbb P(K^{*}=T)
\leq C_{T-1}^{n-2}(1-p_{low})^{T+1-n}p_{up}^{n-1}.
\end{aligned}
\end{equation}
Moreover, the expected coalescence time can be estimated by
\[
(n-1)\frac{p_{low}^{n-1}}{p_{up}^{n}} \leq \mathbb E(K^{*}) \leq (n-1)\frac{p_{up}^{n-1}}{p_{low}^{n}}.
\]
\end{theorem}
{\it Proof.}
 We know that the number of groups will decrease by 1 at the time $k$ if two players coalesce into one group.  Let $\Delta_k$ indicate whether two players coalesce  at time ~$k=1,2,\cdots$ or not, i.e.,
\begin{align*}
\Delta_k=\left\{
\begin{aligned}
&1, \text{players~coalesce at~time}~k,\\
&0, \text{otherwise}.
\end{aligned}
\right.
\end{align*}
By the definition of $\Delta_{k}$ and $K^{*}$, it is easy to know that
\[
\{K^{*}=T\}=\{\Delta_{T}=1,\sum_{k=1}^{T-1}\Delta_{k}=n-2\}.
\]
Consequently,
\begin{equation}\label{eq:P-K-T}
\begin{aligned}
\mathbb P(K^{*}=T)
=\mathbb P(\Delta_{T}=1|\sum_{k=1}^{T-1}\Delta_{k}=n-2)\mathbb P(\sum_{k=1}^{T-1}\Delta_{k}=n-2).
\end{aligned}
\end{equation}
By Lemma \ref{thm:p_low_and_pup},  we have
\[
p_{low}\leq \mathbb  P(\Delta_{k}=1|\Delta_{k-1}=\delta_{k-1},\dots, \Delta_{1}=\delta_{1})\leq p_{up}\]
 and
\[
1-p_{up}\leq \mathbb P(\Delta_{k}=0|\Delta_{k-1}=\delta_{k-1},\dots, \Delta_{1}=\delta_{1})\leq 1-p_{low}
\]
for all $T-1\leq k \leq 1$.
Because
\[
\begin{aligned}
&\mathbb P(\Delta_{T-1}=\delta_{T-1},\dots, \Delta_{1}=\delta_{1})\\
&=\mathbb P(\Delta_{T-1}=\delta_{T-1}|\Delta_{T-2}=\delta_{T-2},\dots, \Delta_{1}=\delta_{1})\cdots \mathbb P(\Delta_{2}=\delta_{2} |\Delta_{1}=\delta_{1})\mathbb P(\Delta_{1}=\delta_{1}),
\end{aligned}
\]
we have
\[
\begin{aligned}
(1-p_{up})^{T+1-n}p_{low}^{n-2} \leq  \mathbb P(\Delta_{T-1}=\delta_{T-1},\dots, \Delta_{1}=\delta_{1})
\leq   (1-p_{low})^{T+1-n}p_{up}^{n-2}
 \end{aligned}
\]
for all $\sum_{k=1}^{T-1} \delta_{k}=n-2$.
Then, it follows from
\[
\begin{aligned}
\mathbb P(\sum_{k=1}^{T-1}\Delta_{k}=n-2)
= \sum_{\sum_{k=1}^{T-1} \delta_{k}=n-2} \mathbb P(\Delta_{T-1}=\delta_{T-1},\dots, \Delta_{1}=\delta_{1})
\end{aligned}
\]
that
\[
\begin{aligned}
C_{T-1}^{n-2}(1-p_{up})^{T+1-n}p_{low}^{n-2}
\leq    \mathbb P(\sum_{k=1}^{T-1}\Delta_{k}=n-2)
\leq  C_{T-1}^{n-2}(1-p_{low})^{T+1-n}p_{up}^{n-2}.
 \end{aligned}
\]
By Lemma \ref{thm:p_low_and_pup},  we have
$$p_{low}\leq \mathbb P(\Delta_{T}=1|\sum_{k=1}^{T-1}\Delta_{k}=n-2)\leq p_{up}.$$
 Therefore, (\ref{eq:estimation-of-K}) holds.

One knows that
\[
\begin{aligned}
 \sum_{T=n-1}^{\infty} TC_{T-1}^{n-2}(1-p_{up})^{T+1-n}p_{low}^{n-1} \leq  \mathbb E(K^{*})
 \leq  \sum_{T=n-1}^{\infty} TC_{T-1}^{n-2}(1-p_{low})^{T+1-n}p_{up}^{n-1}.
\end{aligned}
\]
Denote $s=T-n+1$. It follows from
\[\sum_{T=n-1}^{\infty}C_{T}^{n-1}(1-p_{up})^{T+1-n}=\sum_{s=0}^{\infty}C_{s+n-1}^{s}(1-p_{up})^s=p_{up}^{-n}\]
that
\[
\begin{aligned}
\sum_{T=n-1}^{\infty}TC_{T-1}^{n-2}(1-p_{up})^{T+1-n}p_{low}^{n-1}
 =\sum_{T=n-1}^{\infty}(n-1)C_{T}^{n-1}p_{low}^{n-1}(1-p_{up})^{T+1-n}=(n-1)\frac{p_{low}^{n-1}}{p_{up}^{n}}. \\
\end{aligned}
\]
Similarly, we have
\[
(n-1)\frac{p_{low}^{n-1}}{p_{up}^{n}} \leq \mathbb E(K^{*}) \leq (n-1)\frac{p_{up}^{n-1}}{p_{low}^{n}}.~\blacksquare
\]

\begin{theorem}\label{Consenus}
If Assumptions 1 - 3 hold and all groups interact by playing game $\mathbb G$, then the system reaches coalescence with probability 1.
\end{theorem}
{\it Proof.}
Let $A_k$ be the event that all agents do not coalesce into one group at time $k$. It follows that the event
``all agents do not coalesce into one group''  is $\cup_{l=1}^{\infty} \cup_{k\geq l }^{\infty} A_k$.
 It is easy to find that $A_k=\{\sum_{t=1}^{k}\Delta_{t}<n-1\}$.
By Theorem \ref{thm:est-PK}, we have
\begin{equation*}
\left\{
\begin{aligned}
& \mathbb P(A_k)=1,  k=0,1,\cdots, n-2,\\
&\mathbb P(A_k)=\mathbb P\left(\sum_{t=1}^k\Delta_t<n-1\right)\leq\sum_{s=0}^{n-2}C_k^{s}p_{up}^s(1-p_{low})^{k-s}, \\
  &~~~~~~~~~~~~~ k=n-1, n, \cdots.
\end{aligned}
\right.
\end{equation*}
For $k>n-1$,
\begin{equation}\label{Ak}
\begin{aligned}
\sum_{s=0}^{n-2}C_{k+1}^{s}p_{up}^s(1-p_{low})^{k+1-s}
&=\sum_{s=0}^{n-2}(1-p_{low})\frac{k+1}{k+1-s}C_k^{s}p_{up}^s(1-p_{low})^{k-s}\\
                 &<\frac{(k+1)(1-p_{low})}{k+1-(n-2)}\sum_{s=0}^{n-2}C_k^{s}p_{up}^s(1-p_{low})^{k-s}.
\end{aligned}
\end{equation}
We know that
\begin{equation}\label{ineq-k}
(1-p_{low})\frac{k+1}{k+1-(n-2)}<1-\frac{p_{low}}{2}<1
\end{equation}
holds for all $k>(n-2)(\frac{2}{p_{low}}-1)-1$. Let
$K_{0}=\max\{(n-2)(\frac{2}{p_{low}}-1),n-1\}.$
It follows from (\ref{Ak}) and (\ref{ineq-k}) that
\[
\begin{aligned}
 \sum_{k=K_{0}}^{\infty}\sum_{s=0}^{n-2}C_k^{s}p_{up}^s(1-p_{low})^{k-s}
 &<\sum_{l=0}^{\infty}\sum_{s=0}^{n-2}(1-\frac{p_{low}}{2})^{l}C_{K_{0}}^{s}p_{up}^s(1-p_{low})^{K_{0}-s}\\
 &=\frac{2}{p_{low}}\sum_{s=0}^{n-2}C_{K_{0}}^{s}p_{up}^s(1-p_{low})^{K_{0}-s}<\infty.
 \end{aligned}
 \]
 Thus, we have
\[
\begin{aligned}
\sum_{k=0}^{\infty} \mathbb P(A_k)&=\sum_{k< K_{0}}\mathbb P(A_k)+\sum_{k\geq K_{0}}\mathbb P(A_k)\\
&\leq \sum_{k<K_{0}}\mathbb P(A_k)+\sum_{k=K_{0}}^{\infty}\sum_{s=0}^{n-2}C_k^{s}p_{up}^s(1-p_{low})^{k-s} \\
&<\sum_{k< T}\mathbb P(A_k)+\frac{2}{p_{low}}\sum_{s=0}^{n-2}C_{K_{0}}^{s}p_{up}^s(1-p_{low})^{K_{0}-s}<\infty.
\end{aligned}
\]
By Lemma \ref{Borel¨CCantelli}, we know that
$
\mathbb P(\cup_{l=1}^{\infty} \cup_{k\geq l}^{\infty} A_l)=0,
$
which means that the system will reach coalescence with probability 1. $\blacksquare$


\subsection{Coalescence of multiple agents: special cases}

Generally speaking, $\Delta_{1}, \Delta_{2},\dots,\Delta_{K^{*}}$ are not independent, i.e., the results of game $\mathbb G$ at time $1,2,\dots, k-1$ influence that of at time $k$. However, if $p_{k}$ is independent from $\xi_{k}$, then $\Delta_{1}, \Delta_{2},\dots,\Delta_{K^{*}}$ are independent. We have the following results.

\begin{theorem}\label{thm:special_independent}
If $g(\xi)= \theta \xi^{\lambda}$ and $f(\xi)=c g(\xi)$($\lambda>0, \theta>0,0<c<1$), then
\begin{enumerate}
\item $\Delta_{1}, \Delta_{2},\dots,\Delta_{K^{*}}$  are independent, identically distributed (i.i.d.) random variables, and  the distribution of $\Delta_{k}$ is
\[\begin{aligned}
\mathbb P(\Delta_{k}=1)=\hat p,\mathbb P(\Delta_{k}=0)=\hat q,
\end{aligned}
\]
where $\hat p=1-\left(\frac{c}{2^{\lambda}(1-c)+c}\right)^{2}$ and $\hat q=1-\hat p$;
\item the distribution of $K^*$ is
\begin{equation*}
\mathbb P(K^*=T)=C_{T-1}^{n-2}\hat p^{n-1}\hat q^{T-(n-1)}, ~T=n-1,n,\cdots;
\end{equation*}
\item $\mathbb E(K^*)=\frac{n-1}{\hat p}, \mathbb D(K^*)=\frac{(n-1)\hat q}{\hat p^2}.$
\end{enumerate}
\end{theorem}
\it{Proof.}
  From Corollary \ref{cor1}, we have
  \[p_{k}=1-\left(\frac{f(\frac{\xi_{k}}{2})}{g(\xi_{k})-f(\xi_{k})+f(\frac{\xi_{k}}{2})}\right)^{2}.\]
   Easy to find that $p_{k}=\hat p$, which is independent from $\xi_{k}$. As a result, $\Delta_{1}, \Delta_{2},\dots,\Delta_{K^{*}}$ are independent and $\mathbb P(\Delta_{k}=1)=\hat p$.

  Since $\Delta_1$ $\Delta_2$, $\cdots, \Delta_{K^{*}}$ are i.i.d. random variables. We have
\[
\begin{aligned}
\mathbb P(K^*=T)&=\mathbb P(\Delta_{T}=1,\sum_{i=1}^{T-1}\Delta_{i}=n-2)\\
            &=\mathbb P(\Delta_{T}=1)\mathbb P(\sum_{i=1}^{T-1}\Delta_{i}=n-2)\\
           &=\left\{
           \begin{aligned}
           & 0, && k=0,1,\cdots,n-2,\\
           &C_{T-1}^{n-2}\hat p^{n-1}\hat q^{T-1-(n-2)}, && T=n-1,n,\cdots.
           \end{aligned}
           \right.
\end{aligned}
\]
It follows that the expectation of $K^*$ is
\[
\mathbb E(K^*)=\sum_{T=n-1}^{\infty}TC_{T-1}^{n-2}\hat p^{n-1}\hat q^{T-(n-1)}.
\]
Using the similar argument in Theorem \ref{thm:est-PK}, we have
\[
\mathbb E(K^*)=\frac{n-1}{\hat p}.
\]
 We can also show that
\[
\begin{aligned}
\mathbb E((K^{*})^{2})&=\sum_{T=n-1}^{\infty}T^2C_{T-1}^{n-2}\hat p^{n-1}\hat q^{T-(n-1)}.\\
\end{aligned}
\]
Let $s=T-n+1$, we obtain
\[ \begin{aligned}
\mathbb E((K^{*})^{2})&
              =(n-1)\hat p^{n-1}\sum_{s=0}^{\infty}(s+n-1)C_{s+n-1}^{s}\hat q^{s}\\
           &=(n-1)\hat p^{n-1}\left[n\sum_{s=0}^{\infty}C_{s+n}^{s}\hat q^{s}
           -\sum_{s=0}^{\infty}C_{s+n-1}^{s}\hat q^{s}\right]\\
           &=\frac{n(n-1)}{\hat p^{2}}-\frac{n-1}{\hat p}.
\end{aligned}
\]
Therefore, we have
$
\mathbb D(K^*)=\frac{(n-1)\hat q}{\hat p^2}.
$
$\blacksquare$

%

The expected coalescence time can measure how fast all agents coalesce into one group. By Theorem \ref{thm:special_independent}, we have the following result.

\begin{corollary}\label{co:EK}
If $g(\xi)= \theta \xi^{\lambda}$ and $f(\xi)=c g(\xi)$($\lambda>0, \theta>0,0<c<1$),  then $\mathbb E(K^*)$ is a strictly monotone increasing function of $c$.
\end{corollary}
{\it Proof.} By $\hat p=1-\left(\frac{c}{2^{\lambda}(1-c)+c}\right)^{2}$, we can find that $\hat p$ is a strictly monotonic decreasing function of $c$. And $\mathbb E(K^*)$ is a  strictly monotone decreasing function of $\hat p$. Therefore, $\mathbb E(K^*)$ is a strictly monotone increasing function of $c$. $\blacksquare$

At time $k$, the game $\mathbb G$ is played by two groups with size $s_1$ and $s_2$. When $g(\xi)= \theta \xi^{\lambda}$ and $f(\xi)=c g(\xi)$($\lambda>0, \theta>0,0<c<1$), the aggregate expectational payoff of all agents is
$
 U^{(k)}(p^{*},q^{*})=(s_{1}+s_{2})\frac{2^{\lambda}(1-c)\theta\xi^{\lambda}}{2^{\lambda}(1-c)+c}.
$

\section{Simulations} \label{s4}

\begin{figure}
  \centering
  \includegraphics[width=8.5cm]{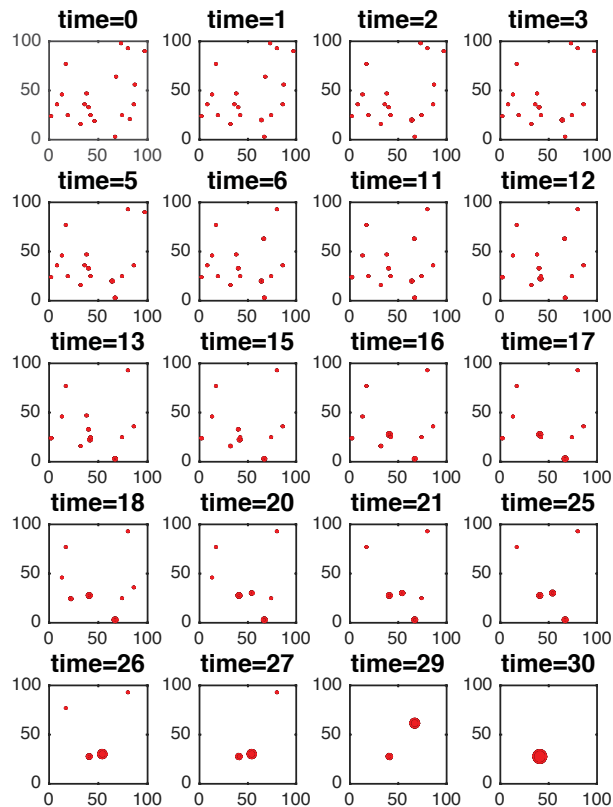}\\
  \caption{The process of coalescing. Each figure indicates states of agents after two groups coalesce into one. Since the states of agents from one group are the same, each dot also indicates one group. We use bigger dots to indicate groups with more agents. It is easy to find that the numbers of groups decreases by 1 at each time. Moreover, some groups become bigger and bigger as time goes by. Finally, all agents coalesce into one big group at time 30.  }\label{fig2}
  \end{figure}

Suppose that there are 20 agents with distinct initial states. Firstly, we let  $g(\xi)=0.8\xi$ and $f(\xi)=\frac{6}{8}g(\xi)$. In Fig.\ref{fig2},  we show the process of coalescing by presenting the groups at time when merging event happens. Since agents from the same group have the same state, each dot indicates one group. In order to show the process clearly, we use bigger dots to indicate groups with more agents.  It is shown that,  when two groups play game $\mathbb G$ and coalesce into a bigger one, the number of groups shrinks by 1. Moreover, some groups become bigger and bigger as time goes by. The system reaches coalescence at time 30.

\begin{figure}
  \centering
  \includegraphics[width=8cm]{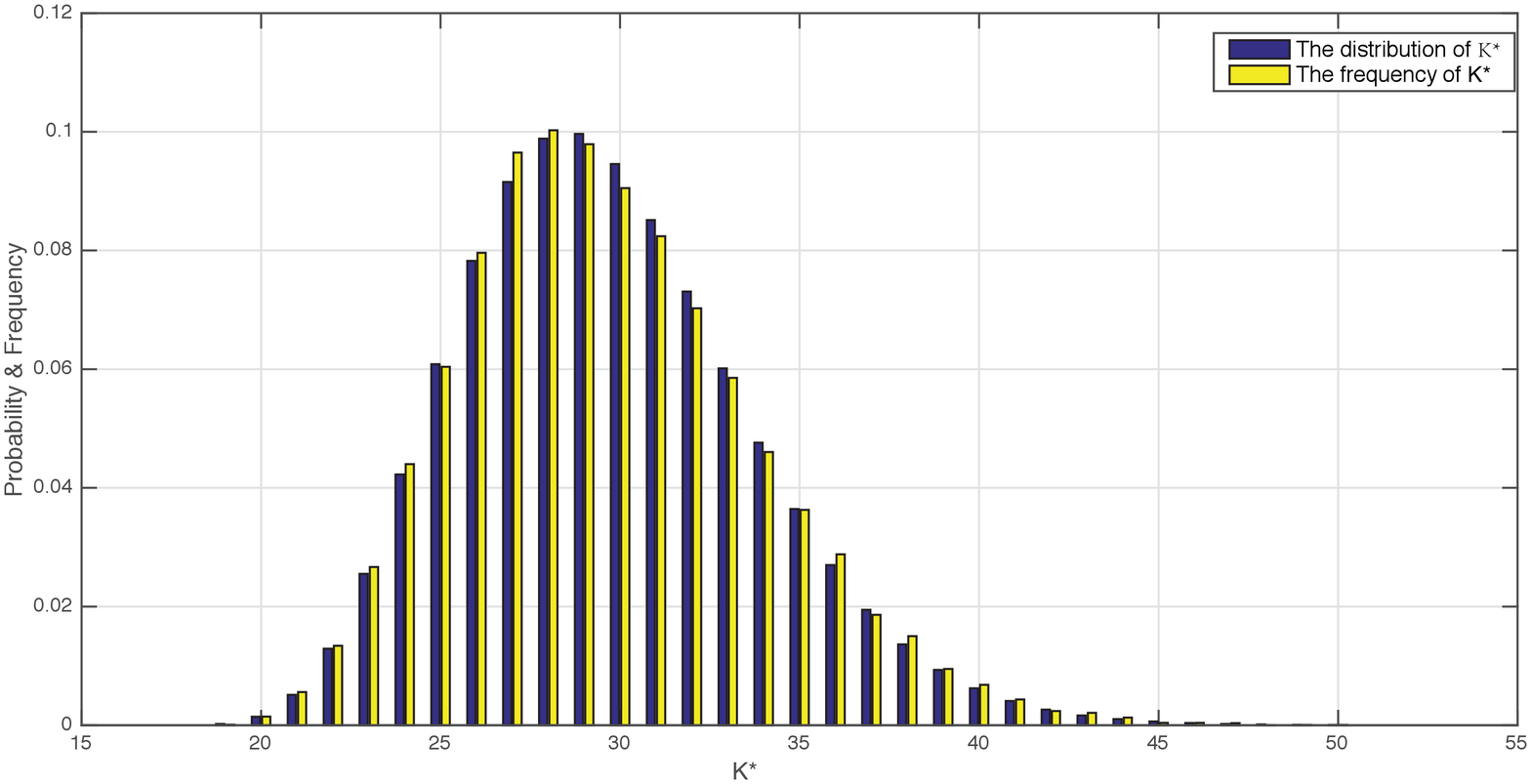}\\
  \caption{The distribution and the frequency of $K^*$ with  $f(\xi)=0.6\xi$ and $g(\xi)=0.8\xi$}\label{fig3}
  \end{figure}

Secondly, We simulate 20000 times with the same initial states. It is shown that each time  the system always achieves coalescence in the finite time. Moreover, we also get the frequency of coalescence time $K^*$ over those 20000 times simulations. The comparison between the distribution and the frequency of $K^*$  is shown in Fig. \ref{fig3}. Those results manifest the effectiveness of theoretical results in Theorems \ref{Consenus} and \ref{thm:special_independent}.

\begin{figure}
  \centering
  \includegraphics[width=7cm]{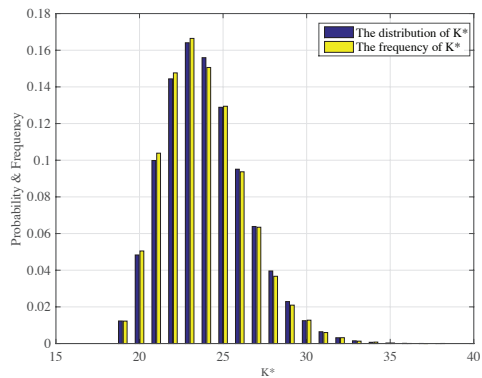}\\
  \caption{The distribution and the frequency of $K^*$ with $f(\xi)=0.5\xi$ and $g(\xi)=0.8\xi$}\label{fig5}
  \end{figure}

Thirdly, we let $g(\xi)=0.8\xi$ and $f(\xi)=\frac{5}{8}g(\xi)$. Then we do the same simulations. The comparison between the distribution and the frequency of $K^*$  is shown in Fig. \ref{fig5}.  Easy to find from Fig. \ref{fig3} and Fig. \ref{fig5} that the system is more likely reaching coalescence earlier when $g(\xi)=0.8\xi$ and $f(\xi)=\frac{5}{8}g(\xi)$. Those results manifest the effectiveness of theoretical results in Corollary \ref{co:EK}.

%



\section{Conclusion}\label{s5}
To achieve some global tasks, multiple agents need to coalesce into one group --- they will make decisions together, share information instantly, keep consensus in states. This paper focused on the coalescence of a population of rational and complete information accessible agents. We modeled the coalescing process as a repeated bimatrix game. Agents form groups and groups coalesce into one bigger group. We proved that coalescence will be reached with probability one and gave an estimation for the expected coalescence time. Moreover, when payoff functions are power functions, the distribution of coalescence time was obtained.  Future work might contain the coalescence under partial information or under  learning mechanisms.




\ifCLASSOPTIONcaptionsoff
  \newpage
\fi



\bibliographystyle{IEEEtran}
\bibliography{texref}

\begin{thebibliography}{10}
\providecommand{\url}[1]{#1}
\csname url@samestyle\endcsname
\providecommand{\newblock}{\relax}
\providecommand{\bibinfo}[2]{#2}
\providecommand{\BIBentrySTDinterwordspacing}{\spaceskip=0pt\relax}
\providecommand{\BIBentryALTinterwordstretchfactor}{4}
\providecommand{\BIBentryALTinterwordspacing}{\spaceskip=\fontdimen2\font plus
\BIBentryALTinterwordstretchfactor\fontdimen3\font minus
  \fontdimen4\font\relax}
\providecommand{\BIBforeignlanguage}[2]{{%
\expandafter\ifx\csname l@#1\endcsname\relax
\typeout{** WARNING: IEEEtran.bst: No hyphenation pattern has been}%
\typeout{** loaded for the language `#1'. Using the pattern for}%
\typeout{** the default language instead.}%
\else
\language=\csname l@#1\endcsname
\fi
#2}}
\providecommand{\BIBdecl}{\relax}
\BIBdecl

\bibitem{Hegselmann2002Opinion}
R.~Hegselmann and U.~Krause, ``Opinion dynamics and bounded confidence models,
  analysis and simulation,'' \emph{Journal of Artificial Societies and Social
  Simulation}, vol.~5, no.~3, pp. 1--33, 2002.

\bibitem{Bullo2017Competitive}
W.~Mei and F.~Bullo, ``Competitive propagation: models, asymptotic behavior and
  quality-seeding games,'' \emph{IEEE Transactions on Network Science and
  Engineering}, vol.~4, no.~2, pp. 83--99, 2017.

\bibitem{du2019evolutionary}
J.~Du, ``An evolutionary game coordinated control approach to division of labor
  in multi-agent systems,'' \emph{IEEE Access}, vol.~7, no.~1, pp.
  124\,295--124\,308, 2019.

\bibitem{kao2014collective}
A.~B. Kao, N.~Miller, C.~Torney, A.~Hartnett, and I.~D. Couzin, ``Collective
  learning and optimal consensus decisions in social animal groups,''
  \emph{PLoS Computational Biology}, vol.~10, no.~8, p. e1003762, 2014.

\bibitem{ren2007distributed}
W.~Ren and E.~Atkins, ``Distributed multi-vehicle coordinated control via local
  information exchange,'' \emph{International Journal of Robust and Nonlinear
  Control}, vol.~17, no. 10-11, pp. 1002--1033, 2007.

\bibitem{ren2005consensus}
W.~Ren and R.~W. Beard, ``Consensus seeking in multiagent systems under
  dynamically changing interaction topologies,'' \emph{IEEE Transactions on
  Automatic Control}, vol.~50, no.~5, pp. 655--661, 2005.

\bibitem{Morin2015Collective}
A.~Morin, J.~B. Caussin, C.~Eloy, and D.~Bartolo, ``Collective motion with
  anticipation: flocking, spinning, and swarming,'' \emph{Physical Review E},
  vol.~91, no.~1, p. e12134, 2015.

\bibitem{zheng2014containment}
Y.~Zheng and L.~Wang, ``Containment control of heterogeneous multi-agent
  systems,'' \emph{International Journal of Control}, vol.~87, no.~1, pp. 1--8,
  2014.

\bibitem{Couzin2005Effective}
I.~D. Couzin, J.~Krause, N.~R. Franks, and S.~A. Levin, ``Effective leadership
  and decision-making in animal groups on the move,'' \emph{Nature}, vol. 433,
  no. 7025, pp. 513--516, 2005.

\bibitem{Pais2014Adaptive}
D.~Pais and N.~E. Leonard, ``Adaptive network dynamics and evolution of
  leadership in collective migration,'' \emph{Physica D: Nonlinear Phenomena},
  vol. 267, no.~2, pp. 81--93, 2014.

\bibitem{Claudio2012dynamics}
C.~Altafini, ``Dynamics of opinion forming in structurally balanced social
  networks,'' \emph{PLoS ONE}, vol.~7, no.~6, p. e38135, 2012.

\bibitem{Jadbabaie2003}
A.~Jadbabaie, J.~Lin, and A.~S. Morse, ``Coordination of groups of mobile
  autonomous agents using nearest neighbor rules,'' \emph{IEEE Transactions on
  Automatic Control}, vol.~48, no.~6, pp. 988--1001, 2003.

\bibitem{Majingying2015LQR}
J.~Ma, Y.~Zheng, and L.~Wang, ``L{QR}-based optimal topology of
  leader-following consensus,'' \emph{International Journal of Robust and
  Nonlinear Control}, vol.~25, no.~17, pp. 3404--3421, 2015.

\bibitem{xie2007consensus}
G.~Xie and L.~Wang, ``Consensus control for a class of networks of dynamic
  agents,'' \emph{International Journal of Robust and Nonlinear Control},
  vol.~17, no. 10-11, pp. 941--959, 2007.

\bibitem{ren2008consensus}
W.~Ren, ``On consensus algorithms for double-integrator dynamics,'' \emph{IEEE
  Transactions on Automatic Control}, vol.~53, no.~6, pp. 1503--1509, 2008.

\bibitem{Qin2017On}
J.~Qin, Q.~Ma, H.~Gao, Y.~Shi, and Y.~Kang, ``On group synchronization for
  interacting clusters of heterogeneous systems,'' \emph{IEEE Transactions on
  Cybernetics}, vol.~47, no.~12, pp. 4122--4133, 2017.

\bibitem{Zheng2017Consensus}
Y.~Zheng, J.~Ma, and L.~Wang, ``Consensus of hybrid multi-agent systems,''
  \emph{IEEE Transactions on Neural Networks and Learning Systems}, vol.~29,
  no.~4, pp. 1359--1365, 2018.

\bibitem{ma2019consensus}
J.~Ma, M.~Ye, Y.~Zheng, and Y.~Zhu, ``Consensus analysis of hybrid multiagent
  systems: A game-theoretic approach,'' \emph{International Journal of Robust
  and Nonlinear Control}, vol.~29, no.~6, pp. 1840--1853, 2019.

\bibitem{ma2016Equilibrium}
J.~Ma, Y.~Zheng, B.~Wu, and L.~Wang, ``Equilibrium topology of multi-agent
  systems with two leaders: a zero-sum game perspective,'' \emph{Automatica},
  vol.~73, no.~C, pp. 200--206, 2016.

\bibitem{jackson2002evolution}
M.~O. Jackson and A.~Watts, ``The evolution of social and economic networks,''
  \emph{Journal of Economic Theory}, vol. 106, no.~2, pp. 265--295, 2002.

\bibitem{cox1989coalescing}
J.~T. Cox, ``Coalescing random walks and voter model consensus times on the
  torus in $\mathbb z^d$,'' \emph{The Annals of Probability}, vol.~17, no.~4,
  pp. 1333--1366, 1989.

\bibitem{poduri2007latency}
S.~Poduri and G.~S. Sukhatme, ``Latency analysis of coalescence for robot
  groups,'' in \emph{Proceedings of 2007 IEEE International Conference on
  Robotics and Automation}.\hskip 1em plus 0.5em minus 0.4em\relax IEEE, 2007,
  pp. 3295--3300.

\bibitem{cooper2013coalescing}
C.~Cooper, R.~Els$\ddot{a}$sser, H.~Ono, and T.~Radzik, ``Coalescing random
  walks and voting on connected graphs,'' \emph{SIAM Journal on Discrete
  Mathematics}, vol.~27, no.~4, pp. 1748--1758, 2013.

\bibitem{Ba1999Dynamic}
T.~Ba{\c{S}}ar and G.~J. Olsder, \emph{Dynamic Noncooperative Game Theory, 2nd
  ed.}\hskip 1em plus 0.5em minus 0.4em\relax Academic: San Diego, 1999.

\bibitem{Gut2005Probability}
A.~Gut, \emph{Probability: A Graduate Course}.\hskip 1em plus 0.5em minus
  0.4em\relax Springer: New York, 2005.

\end{thebibliography}

\end{document}